\documentclass[conference]{IEEEtran}
\IEEEoverridecommandlockouts
\usepackage{cite}
\usepackage{amsmath,amssymb,amsfonts}
\usepackage{algorithmic}
\usepackage{graphicx}
\usepackage{textcomp}
\usepackage{xcolor}
\usepackage{caption}
\usepackage{subcaption}
\usepackage[ruled]{algorithm2e}
\usepackage{caption}
\usepackage{subcaption}
\usepackage{epstopdf}
\usepackage{multirow}
\usepackage{svg}
\usepackage{float}

\newcommand{\real}[1]{\mathcal{R} \left \{ #1 \right \}}
\newcommand{\imag}[1]{\mathcal{I} \left \{ #1 \right \}}

\newcommand{\sss}{\mathcal{S}}
\newcommand{\Dtr}{\mathcal{D}_{tr}}
\newcommand{\Dtest}{\mathcal{D}_{test}}

\def\BibTeX{{\rm B\kern-.05em{\sc i\kern-.025em b}\kern-.08em
    T\kern-.1667em\lower.7ex\hbox{E}\kern-.125emX}}
\begin{document}
% A Physics-Augmented Adaptation Method for Deep Learning-Based Underwater Acoustic Source Ranging
\title{Joint Source-Environment Adaptation for Deep Learning-Based Underwater Acoustic Source Ranging
\thanks{This work has been supported by the Office of Naval Research (ONR) under grant N00014-19-1-2662.}}

\author{\IEEEauthorblockN{Dariush Kari}
\IEEEauthorblockA{
\textit{University of Illinois Urbana-Champaign}\\
Urbana, IL, USA \\
dkari2@illinois.edu}
\and 
\IEEEauthorblockN{Andrew C. Singer}
\IEEEauthorblockA{
\textit{Stony Brook University}\\
Stony Brook, NY, USA \\
andrew.c.singer@stonybrook.edu}
}

\maketitle

\begin{abstract}
In this paper, we propose a method to adapt a pre-trained deep-learning-based model for underwater acoustic localization to a new environment. We use unsupervised domain adaptation to improve the generalization performance of the model, i.e., using an unsupervised loss, fine-tune the pre-trained network parameters without access to any labels of the target environment or any data used to pre-train the model. This method improves the pre-trained model prediction by coupling that with an almost independent estimation based on the received signal energy (that depends on the source). We show the effectiveness of this approach on Bellhop generated data in an environment similar to that of the SWellEx-96 experiment contaminated with real ocean noise from the KAM11 experiment.
\end{abstract}

% {This method enables us to effectively incorporate a priori knowledge about the set of the source locations into the model}


\section{Introduction}
% Deep-learning-based (DL) algorithms for underwater acoustic (UWA) localization \cite{niu2017source, wang2019deep, yangzhou2019deep, chen2021model, wang2018underwater} usually generalize poorly to new environments \cite{liu2023unsupervised} due to environmental mismatches, although they liberate us from the need to have a priori knowledge about the environment, as they inherently learn the necessary parameters through training. 
% Given the complete information about an environment, matched field processing (MFP) \cite{baggeroer1988matched,sullivan1993estimation} provides a reasonable solution to underwater acoustic (UWA) localization, however, it is prone to environmental mismatches such as the mismatches in sound speed profile or bathymetry. The absence of accurate environmental knowledge encourages us to use deep learning  (DL)-based methods. 
Despite the success of deep learning (DL)-based underwater acoustic (UWA) localization algorithms \cite{niu2017source, chen2021model, wang2018underwater}, their applications remain limited due to poor generalization in mismatched environments\cite{chen2021model, liu2023unsupervised}. Domain adaptation (DA) methods can mitigate small mismatches between testing and training datasets by aligning their statistics \cite{ganin2015unsupervised}, but these approaches require access to training data. However, with the growing adoption of decentralized processing \cite{song2020underwater, zhuo2020auv, zhang2020node}, training data will not always be available due to the communication cost of transferring large datasets between low-power underwater devices or due to security or privacy issues. Therefore, this paper seeks to improve the generalization performance of DL-based UWA localization models via \emph{source-free} DA \cite{liang2020we}.\par

In an effort to provide a more robust solution than matched field processing (MFP) \cite{baggeroer1988matched,sullivan1993estimation}, Chen and Schmidt \cite{chen2021model} propose a model-based convolutional neural network (CNN), which outperforms MFP in the presence of depth mismatch. To generate the training dataset, the authors use OASES \cite{schmidt2012ocean} with the sound speed profile set according to the values measured in the real experiment. While a reasonable way to generate more realistic training data, this method does not consider the parameter shift between the synthetic training and the real testing environments. Moreover, it assumes a prior knowledge of the parameters of the test environment, which is rarely available. This motivates further improvement using an adaptation mechanism.\par

The environmental mismatch problem is known in the machine learning literature as domain shift \cite{zhang2013domain} and there are different approaches to its solution, including domain adaptation or domain generalization \cite{zhang2013domain, liang2020we}, transfer learning \cite{wang2019deep, yosinski2014transferable}, and data augmentation \cite{yao2022improving}. While data augmentation and transfer learning have been studied for some cases in underwater acoustic localization \cite{wang2019deep, liu2021deep}, DA remains mainly unexplored, with sparse studies like \cite{liu2023unsupervised} and \cite{long2023deep}.

%Moreover, transfer learning requires labels of some of the test data, which is usually not available in UWA problems and encourages pursuit of DA methods. \par

% {\color{red} cite Amir's paper in this part.} As an example of the supervised machine learning algorithms in UWA localization, \cite{niu2017source} implemented and compared the localization performance (both as a classification and a regression problem) of a feedforward neural network, a support vector machine, and a random forest, where they use the normalized sample covariance matrix from a vertical linear array as the input to these algorithms. Additionally, \cite{yangzhou2019deep} provides successful implementation of a deep neural network localization in a shallow water tank experiments.

% problem of DL-based: data scarcity --> hence, transfer learning \cite{wang2019deep} or domain adaptation. {\color{red}"Deep transfer learning (DTL) \cite{yosinski2014transferable} migrates
% the predictive ability of a trained DNN into a new similar environment through sharing some DNN parameters and rediscovering others. This paper adapts DTL for
% source ranging in an uninvestigated deep-sea area, which takes full advantage of the
% historical environment data and limited experimental acoustic data."}  
Whenever accessible, DA can leverage training data to learn the domain shift. Nonetheless, source-free DA \cite{liang2020we} is still possible by inferring some useful statistics (features) of the training set from the pre-trained model. Note that not only can source-free DA respect the privacy in distributed settings, but also it can prevent the communication cost tied to large data-set transfer in a challenging underwater environment. 
% In such scenarios, instead of dependence on training data, the adaptation process assumes prior knowledge about the target data distribution and the domain shift.\par
% Specifically, in this paper, we assume that test data form clear clusters of data and are almost evenly distributed over the area of interest. Based on this assumption, we encourage the network to generate more diverse outputs by maximizing the \emph{outputs average} entropy, similar to \cite{liang2020we} and \cite{ding2022source}.\par

% {\color{red} the importance of pseudo-labels, based on the model uncertainty -- hence the necessity of obtaining uncertainty.} 
As a source-free DA method, source hypothesis transfer (SHOT) \cite{liang2020we} assumes that the DL-based model consists of a domain-specific feature extractor and a domain-invariant classifier (hypothesis). During the inference, while freezing the classifier, SHOT fine-tunes the feature extractor to produce confident outputs and prevents it from a degenerated solution by using an information maximization loss \cite{liang2020we} that promotes diverse output labels. While this procedure is useful for one-hot coded labels, it is not as effective for cases where labels are not equidistant, e.g., for localization.\par
% {\color{red}Note that the success of this method relies on the validity of the assumption about the distribution shift.}

% To accomplish a source-free DA for a DL-based UWA localization network, we equip a pre-trained CNN with an adaptation loss that consists of two parts: a self-supervised loss that encourages the model to make more confident predictions on samples for which the network is believed to have a good performance, and an entropy maximization part to prevent all predicted labels from degenerating to those of the self-supervising samples. To encourage more confident predictions, we need some notion of model uncertainty about its predictions. 

% {\color{red} discuss the MC dropout and its relation to uncertainty, hence motivate the use of classification network. Also, how is this related to conformal prediction?}
Although localization naturally fits into a regression paradigm, regression in its standard form only outputs a point estimate, while a classification model provides a probability mass function (PMF). Hence, we use a classification model to obtain the model uncertainty from the output PMF, e.g., a unimodal PMF implies the model is more certain about its prediction compared to the case of a multimodal PMF. In addition, by using a metric-inspired label softening to preserve distances between different classes, we make the classification model more amenable to the localization problem.\par

\begin{figure}[htb]
\centering
    \includegraphics[width= 0.7\linewidth]{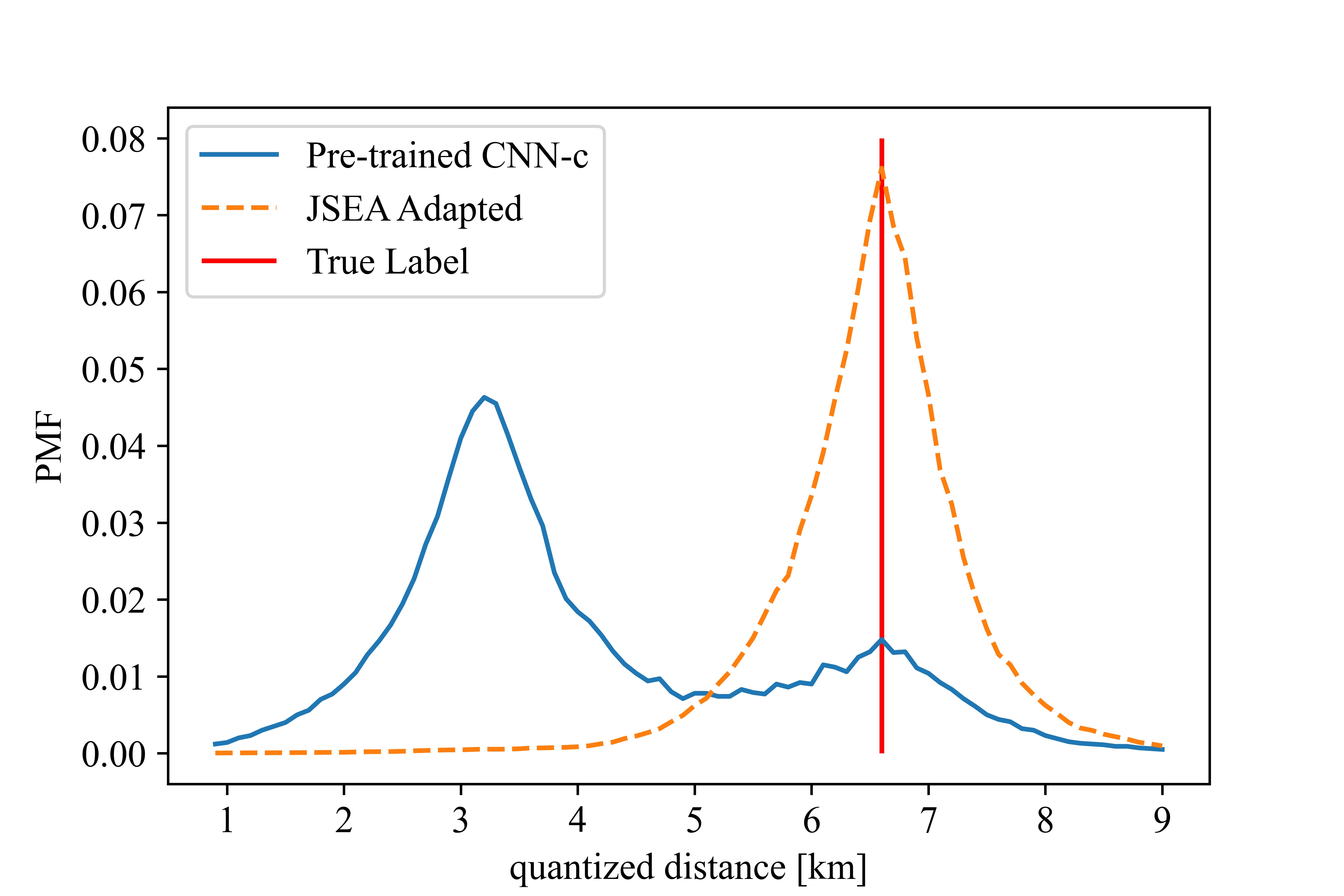}
    \caption{An example of a predicted label in a test environment with a different sound speed profile than the training environment. Observe that the pre-trained model prediction (the label with maximum probability) is far from the ground truth, however, there is another peak in the output closer to the ground-truth.}
    \label{fig:label_example}
\end{figure}

Figure \ref{fig:label_example} shows an example output of a pre-trained classification model under environmental mismatch, where the largest peak in the output is not necessarily aligned with the true label. To decide which peak to select, we leverage another estimation process based on the received signal strength, which is potentially more robust and is only required to provide a coarse estimate. Note that array-based localization usually involves normalization by the total received signal power at the array elements, which is supposed to remove the source power information and make the algorithm robust to the source power. There are also methods that use the received signal strength for localization \cite{xu2016rss, zhang2016received}. We propose that combining these two approaches in a Bayesian manner can enhance the generalization performance.

\section{Localization Using Sample Covariance Matrices}
\textbf{Notation:} Bold letters denote matrices and vectors, $H(\mathbf{y})$ indicates the discrete entropy of the PMF $\mathbf{y}$, $D_{KL}$ denotes the KL-divergence, $\real{x}$ and $\imag{x}$ respectively denote the real and imaginary parts of $x$, $\mathbf{A}^{\mathsf{T}}$ and $\mathbf{A}^{\mathsf{H}}$ denote respectively the transpose and conjugate transpose of the matrix or vector $\mathbf{A}$. Also, for a set of test points $\sss$, the complement is shown by $\sss^\mathsf{c} = \Dtest - \sss$ and the cardinality is shown by $|\sss|$.\par

Sample covariance matrices (SCM) are usually used for UWA localization using arrays because they are sufficient statistics for the jointly Gaussian signal and noise \cite{gerstoft2020parametric} scenarios. The normalized SCM of an $L$ element array using $P$ snapshots $\mathbf{\Tilde{R}}^{(p)}(f)$ is calculated as 
\begin{equation}\label{eq:SCM}
\tilde{\mathbf{C}} = \frac 1 P \sum_{p=1}^{P} \mathbf{\Tilde{R}}^{(p)}(f) \mathbf{\Tilde{R}}^{(p)}(f)^{\mathsf{H}},
\end{equation}
where $\mathbf{\Tilde{R}}^{(p)}(f) = [ \Tilde{R}^{(p)}_1(f),  \Tilde{R}^{(p)}_2(f), ...,  \Tilde{R}^{(p)}_L(f)]^\mathsf{T}$ and
\begin{align}
    \Tilde{R}^{(p)}_l(f) &= \frac{R^{(p)}_l(f)}{\sqrt{\sum_{l=1}^{L} |R_l^{(p)}(f)|^2}}, \quad l \in \{1,2,...,L \},
\end{align}
where the scalar $R^{(p)}_l(f)$ is the complex Fourier coefficient of the $p$-th segment of the $l$-th hydrophone received signal at the frequency $f$. We use a $3$-second-long signal divided into $P=5$ segments, each of length $1$ second and a $50\%$ overlap, tapered by a Kaiser window with $\beta=9.24$. The goal is to determine the source range $d$, given the corresponding SCM $\mathbf{C}$.
Since we assume that the source signal is monotone at frequency $f$, we will drop $f$ from the notation in the following sections.\par

\subsection{MFP approach}
We use the Bartlett processor \cite{gemba2017adaptive} as an MFP baseline, which is described by
\begin{equation}
    \hat{d} = \arg \max_{d} \mathbf{\Tilde{R}}(d)^{\mathsf{H}} \Tilde{\mathbf{C}} \mathbf{\Tilde{R}}(d),
\end{equation}
where $\Tilde{\mathbf{C}}$ is the normalized SCM of the measured data and $\mathbf{\Tilde{R}}(d)$ is the replica field generated by a source at range $d$. For each $d$, the corresponding $\mathbf{\Tilde{R}}(d)$ from the training set is used at the inference time. Observe that this method has access to training data at the test time, however, we do not adapt $\mathbf{\Tilde{R}}(d)$ to the new environment.

\subsection{CNN approach}
\begin{figure}[htb]
\centering
    \includegraphics[width= 0.7\linewidth]{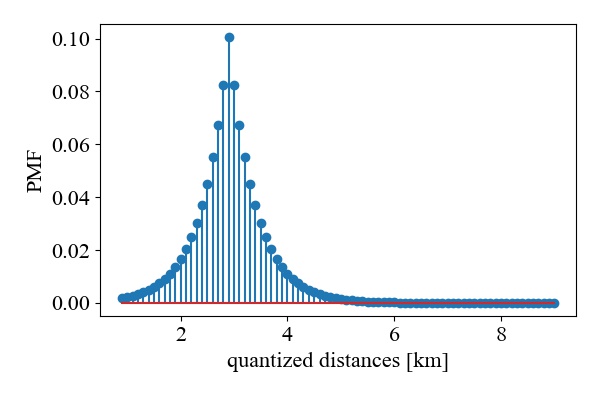}
    \caption{An example of a metric-inspired smoothed label $y_i$ corresponding to $d^q_i = 20$. Here, we have used the absolute error as the metric, hence, we have used a truncated exponential as the label.}
    \label{fig:smoothed}
\end{figure}

As depicted in Fig. \ref{fig:sfda}, the feature extractor part of the model is a CNN followed by a linear layer, that takes the real and imaginary parts of the SCM, $[\real{\Tilde{\mathbf{C}_i}} \, | \, \imag{\Tilde{\mathbf{C}_i}}]$, as the $2$-channel input (a real tensor of dimension $2 \times L \times L$) and generates the feature vector $\boldsymbol{\phi}_i$. These features are then fed into a linear classifier to generate the output PMF $\hat{\mathbf{y}}_i$. Denoting the area of interest for the source range by $[D_{\min}, \; D_{\max}]$, for each training sample $i$, we quantize the range $d_i$ by $d^q_i = \lfloor \frac{d_i - D_{\min}}{100} + 0.5 \rfloor$. Although in classification tasks, usually a one-hot representation of $d^q_i$ is used, since we use mean absolute error (MAE) to evaluate our results, we define the soft label $ \mathbf{y}_i = [y_{i1}, y_{i2}, ..., y_{iM}] = \eta(d^q_i)$ as
\begin{align}
    y_{ik} = \frac {\exp (-|k - d^q_i|/\sigma)}{\sum_{k=1}^M \exp (-|k - d^q_i|/\sigma)},
    % A_i = \sum_{k=1}^M \exp (-|k - g_i|/\sigma),
\end{align}
where, $M$ is the number of classes and $\sigma$ is a hyper-parameter. Figure \ref{fig:smoothed} shows one such smoothed label.

Therefore, the training dataset is denoted by $\Dtr = \{([\real{\Tilde{\mathbf{C}_i}} \, | \, \imag{\Tilde{\mathbf{C}_i}}], \mathbf{y}_i)\}_{i=1}^{N_{tr}}$ and the test set that does not include labels is denoted by $\Dtest = \{[\real{\tilde{\mathbf{C}_i}} \, | \, \imag{\Tilde{\mathbf{C}_i}}]\}_{i=1}^{N_{test}}$. We use the loss function $\mathcal{L}_{tr} = \frac 1 {N_{tr}} \sum_{i=1}^{N_{tr}} \mathsf{JSD}(\mathbf{y}_i, \hat{\mathbf{y}}_i)$ for training, where $\hat{\mathbf{y}}_i$ is the network output and $\mathsf{JSD}$ is the specific Jensen-Shannon Divergence \cite{fuglede2004jensen} defined by
\begin{equation}
    \mathsf{JSD}(\mathbf{y}_i, \hat{\mathbf{y}}_i) = \frac 1 2 \left(D_{KL}(\mathbf{y}_i\|\bar{\mathbf{y}}_i) + D_{KL}(\hat{\mathbf{y}}_i\|\bar{\mathbf{y}}_i) \right),
\end{equation}
where $\bar{\mathbf{y}}_i = (\mathbf{y}_i + \hat{\mathbf{y}}_i)/2$.

The CNN structure is similar to that in \cite{chen2021model} with a $2$ channel input, $3$ convolutional layers with $6$, $38$, and $40$ channels and kernels of size $3$, $5$, $5$, respectively. The output of the feature extraction is a $256$-dimensional vector $\boldsymbol{\phi}_i$, which is then fed into the classifier to generate $\hat{\mathbf{y}}_i$.

\section{Domain Adaptation}
\subsection{Environment Adaptation Using SHOT}
When presented with data from a different environment, the network makes \emph{uncertain} predictions, i.e., it will be confused between different classes, which leads to inaccurate predictions. However, since there is no labeled data from the new environment, we cannot supervise the model to make more accurate predictions. Nevertheless, we can encourage the model to preserve its predictions, with a higher confidence, on samples that match to the training data. To this end, we select a self-supervising subset $\sss$ of the samples and use the pre-trained network's output on these samples as their pseudo-labels, for which we use a loss term $\mathsf{JSD}(\mathbf{y}_i^{\mathsf{pseudo}}, \hat{\mathbf{y}}_i)$. Here, $\mathbf{y}_i^{\mathsf{pseudo}}$ is the softened labeled based on the output of the pre-trained classifier. However, to prevent the network from only predicting classes that exist in the self-supervising subset, for all of the samples, we add a loss term that seeks to maximize the entropy of the outputs average $\bar{\mathbf{y}} = \frac 1 {N_{test}} \sum_{i=1}^{N_{test}} \hat{\mathbf{y}}_i$.\par

\begin{figure}[ht]
\centering
    \includegraphics[width= 0.7\linewidth]{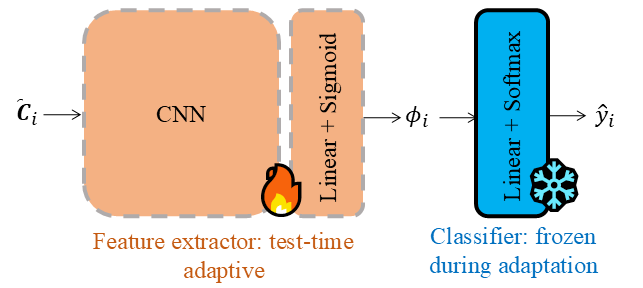}
    \caption{The CNN classifier includes a feature extraction part followed by a linear classifier that will be frozen during adaptation.}
    \label{fig:sfda}
\end{figure}

To determine the self supervising set $\sss$, first note that we freeze the pre-trained classifier as depicted in Fig. \ref{fig:sfda}, which predicts the label $\hat{\mathbf{y}}_i$ by
\begin{equation}
    \hat{\mathbf{y}}_i = \text{Softmax}(\mathbf{w}_k^\mathsf{T} \boldsymbol{\phi}_i),
\end{equation}
where $\mathbf{w}_k$ is the classification weight vector corresponding to the $k$-th class. Note that if the feature vector $\boldsymbol{\phi}_i$ is highly aligned with $\mathbf{w}_{k_0}$ for one $k_0$, then the network predicts a unimodal soft label $\hat{\mathbf{y}}_i$, as opposed to situations where the network generates a multimodal output and is not certain about the predicted label. Inspired by this observation, we construct the self-supervising subset $\sss$ according to
\begin{equation} \label{eq:SSS}
\sss = \{i \; : 1 \leq i \leq N_{test} \, \, \text{s.t.} \, \, \hat{\mathbf{y}}_i \text{ has only 1 significant peak} \},
\end{equation}
% \begin{equation} \label{eq:SSS}
% \sss = \{i \; : 1 \leq i \leq N_{test} \quad \text{s.t.} \quad \min_k \mathsf{cd} (\boldsymbol{\phi}_i, \mathbf{c}^{(5)}_k) \leq \tau \}.
% \end{equation}
where a significant peak is either the only peak or it is more than $10$ times larger than the second largest peak.
\par

% Domain adaptation can be performed by the prior assumption that the test samples create clusters.
% by using an information maximization loss \cite{}, one can update the weights of a pre-trained model.

% Hence, we use the information maximization loss \cite{liang2020we} for the unsupervised adaptation, i.e.,
To perform source hypothesis, we define the SHOT loss by
% \begin{align}\label{eq:DA_loss}
%     \mathcal{L}_{DA} &= - H \left( \bar{\mathbf{y}} \right) + \frac \beta {|\sss|} \sum_{i \in \sss} \mathcal{L}_{KL}(\mathbf{y}_i^{\mathsf{fr}}, \hat{\mathbf{y}}_i) \nonumber \\
%     & \quad + \frac \beta {|\sss^{\mathsf{c}}|} \sum_{i \in \sss^{\mathsf{c}}} \mathcal{L}_{KL}(\eta(\arg\max(\hat{\mathbf{y}}_i)), \hat{\mathbf{y}}_i),
% \end{align}
\begin{align}\label{eq:SHOT_loss}
    \mathcal{L}_{\text{SHOT}} &= - H \left( \bar{\mathbf{y}} \right) + \frac \beta {|\sss|} \sum_{i \in \sss} \mathsf{JSD}(\mathbf{y}_i^{\mathsf{pseudo}}, \hat{\mathbf{y}}_i),
\end{align}
% \begin{equation}
%     \mathcal{L}_{DA} = - H \left(\frac 1 {N_{test}} \sum_{i=1}^{N_{test}} \hat{y}_i \right) + \frac 1 {N_{test}} \sum_{i=1}^{N_{test}} H \left(\hat{y}_i \right),
% \end{equation}
where, $\beta$ is a positive weight. We then employ the Adam optimizer \cite{Kingma2015Adam} to minimize $\mathcal{L}_{\text{SHOT}}$ with a learning rate of $\mu_{DA}$. This loss encourages the network to adapt itself in a way to be more certain about its predictions, while preventing it from assigning all inputs to the same output by encouraging diverse outputs \cite{liang2020we}. This approach is effective for rectifying errors due to small mismatches.

\subsection{Joint Source-Environment Adaptation (JSEA)}
Given that $\mathbf{C} = E \; \tilde{\mathbf{C}}$, where $E = \sum_{l=1}^{L} |R_l^{(p)}(f)|^2$ (which is accurate for $P=1$ snapshot case and approximately correct for $P>1$), and assuming that $E$ and $\tilde{\mathbf{C}}$ are independent,
\begin{align}
    p(\mathbf{y} | \mathbf{C}) &= p(\mathbf{y} | \tilde{\mathbf{C}}, E) = \frac{p(\tilde{\mathbf{C}}, E | \mathbf{y}) p(\mathbf{y})}{p(\tilde{\mathbf{C}}) p(E)} \nonumber \\
    &= \frac{p(\tilde{\mathbf{C}} | E, \mathbf{y}) p(E|\mathbf{y}) p(\mathbf{y})}{p(\tilde{\mathbf{C}}) p(E)} = \frac{ p(\mathbf{y} | \tilde{\mathbf{C}}) p(\mathbf{y} | E)}{p(\mathbf{y})}.
\end{align}
As a result, if $\mathbf{y}$ is uniformly distributed over the area of interest, we have $p(\mathbf{y} | \mathbf{C}) \propto p(\mathbf{y} | \tilde{\mathbf{C}}) p(\mathbf{y} | E)$.
% \begin{equation}\label{eq:product}
%     p(\mathbf{y} | \mathbf{C}) \propto p(\mathbf{y} | \tilde{\mathbf{C}}) p(\mathbf{y} | E).
% \end{equation}
$E$ depends on the source power that can be estimated from $\sss$ in \eqref{eq:SSS}. Note that both $p(\mathbf{y} | \tilde{\mathbf{C}})$ and $p(\mathbf{y} | E)$ can be used for localization. However, $p(\mathbf{y} | \tilde{\mathbf{C}})$ provides a finer estimate (and typically more sensitive to mismatches), while $p(\mathbf{y} | E)$ provides a coarser estimate that is more robust to environmental mismatches. Therefore, by exploiting both terms we can significantly improve the performance of localization algorithms in the presence of mismatches. \par

% {\color{red} put a figure to show robustness of $p(\mathbf{y} | E)$}\par

To achieve $p(\mathbf{y}|E)$, we assume that $E | d \sim \mathcal{N}(\Gamma_s(d), \sigma_s^2)$,
% \begin{equation}
%     E | d \sim \mathcal{N}(\Gamma_s(d), \sigma_s^2),
% \end{equation}
where $\Gamma_s$ (transmission loss) is a source-environment dependent decreasing function of $d$, and $\sigma_s$ is an environmental dependent hyper-parameter. Note that although this assumption is not accurate (since $E \geq 0$), the result is still useful for our purpose. For simplicity, we assume $\Gamma$ is piece-wise constant, hence, can be estimated from the self-supervised samples by $\hat{\Gamma}_{\delta}(d) = \frac{1}{|\sss_{\delta}(d)|} \sum_{k \in \sss_{\delta}(d)} E_k$,
% \[
% \hat{\Gamma}_{\delta}(d) = \frac{1}{|\sss_{\delta}(d)|} \sum_{k \in \sss_{\delta}(d)} E_k,
% \]
where $\sss_{\delta}(d) = \{k \in \sss; |d-\hat{d}_k| \leq \delta \}$, and $E_k$ is the received energy of the $k$-th sample. Then for each $i \in \sss^c$ and $k=1, ..., M$, $p(E_i|d^q_k) \propto \exp (-\frac{|E_i - \hat{\Gamma}_{\delta}(d_k^q)|^2} {2 \sigma_s^2})$,
% \begin{equation}
%     p(E_i|d^q_k) \propto \exp (-\frac{|E_i - \hat{\Gamma}_{\delta}(d_k^q)|^2} {2 \sigma_s^2}),
% \end{equation}
which can be then converted to $p(\mathbf{y}|E)$ using $p(E_i) = \sum_{\mathbf{y}} p(E_i|\mathbf{y}) p(\mathbf{y})$ and $p(\mathbf{y}) = 1 / M$.\par

However, since we only need a coarse estimate from the source adaptation part, to rely more on the neural network for the finer estimation, we obtain the pseudo label by $\mathbf{y}_i^{\mathsf{pseudo}} = \eta(d^q_{k^*})$, where
\begin{equation} \label{eq:pseudo}
    k^* = \arg\min_{k \in \mathcal{P}(i)} |E_i - \hat{\Gamma}_{\delta}(d_k^q)|,
\end{equation}
and $\mathcal{P}(i)$ denotes the set of output peaks for sample $i$. Having defined the pseudo-labels for both $\sss$ and $\sss^c$, we finally adapt the feature extractor using the JSEA loss defined by

\begin{align}\label{eq:JSEA_loss}
    \mathcal{L}_{\text{JSEA}} &= \sum_{i=1}^{N_{test}} \mathsf{JSD}(\mathbf{y}_i^{\mathsf{pseudo}}, \hat{\mathbf{y}}_i).
\end{align}

\section{Empirical Evaluation} \label{sec:simulations}
\begin{figure}[ht]
\centering
    \includegraphics[width=0.7\linewidth]{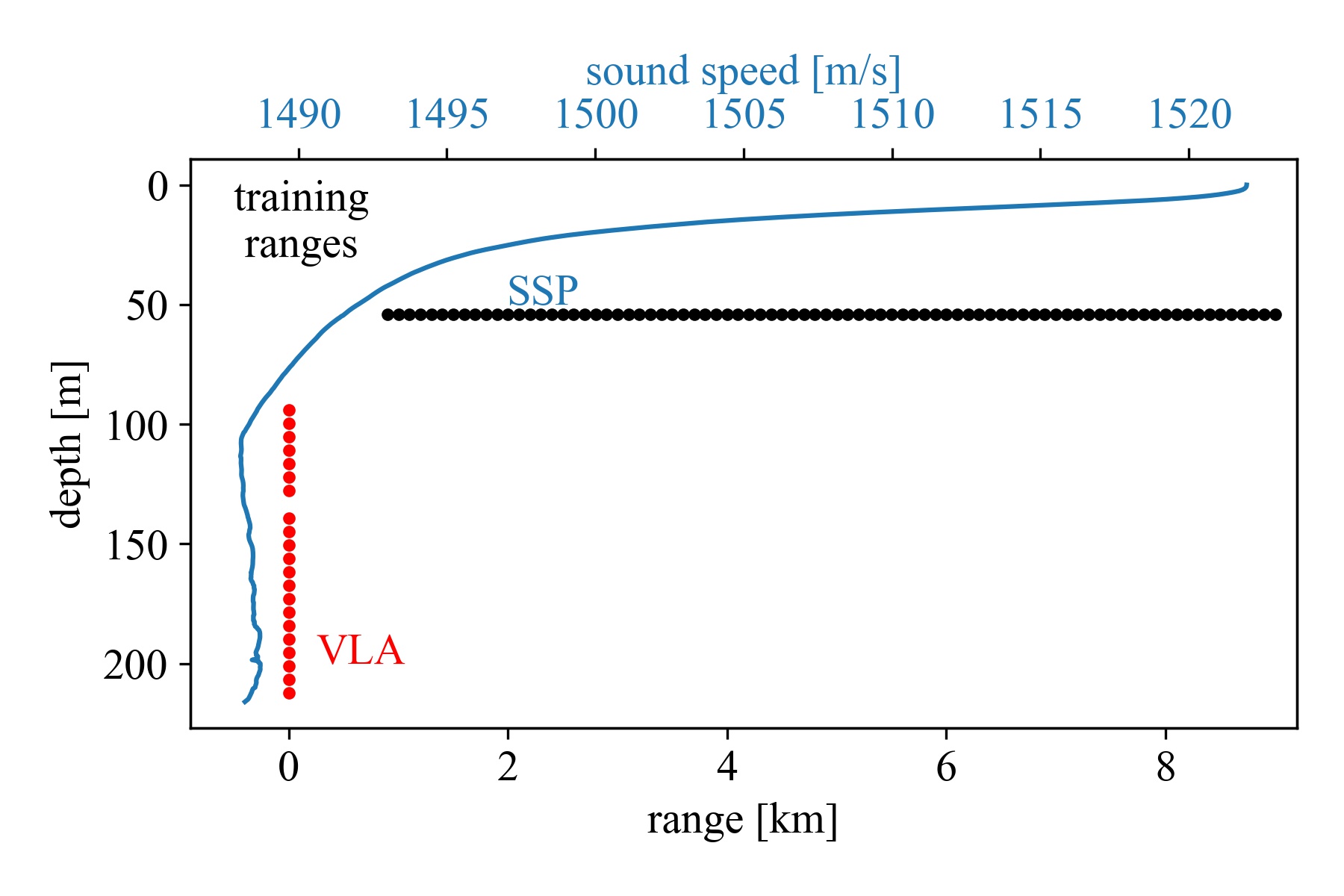}
    \caption{Bellhop environment inspired by the SWellEx-96.}
    \label{fig:swellex}
\end{figure}

We generate data using Bellhop \cite{porter2011bellhop}, where the environmental parameters are set according to the SWellEx-96 \cite{gemba2017adaptive,wang2018underwater} experiment. As depicted in Fig. \ref{fig:swellex}, the training data consist of source ranges between $900$ m and $9$ km with $10$ m increments (thus, $811$ distinct samples), and a $21$-element vertical linear array (VLA) placed at depths from $94.125$ m to $212.25$ m to mimic the SWellEx-96 data. Moreover, the source depth is $54$ m and it transmits a monotone signal with frequency $f=130$ Hz. Note that we have considered a single-layered bottom with a flat bathymetry at depth $216$ m in our simulations and used the average sound speed profile from the SWellEx-96 measurements.\par
% In addition, although we use Gaussian white noise in the training phase to make the model more robust against noise, to emulate the real ocean environment in the test time, we inject noise recorded during the KAM11 \cite{tomasi2011predictability} experiment.\par

To generate a test environment that has a sound speed $c(z)$ at depth $z$, slightly different from that of the training environment, $c_0(z)$, we use a constant gradient perturbation as $c(z) = c_0(z) + \frac{\Delta c}{215} (z - 215)$, where $0 \leq z \leq 215$. We generate data according to random source ranges between $D_{\min} = 900$ m and $D_{\max} = 9$ km, and divide the region into $100$ m-long intervals, each represented by its mid point. Consequently, the labels and classifier outputs are $M=82$ dimensional. In addition to emulating the real ocean environment in the test time, we add a randomly selected noise segment $w$ from the KAM11 experiment \cite{tomasi2011predictability} (scaled to have a desired noise power) to the generated signal for each hydrophone. Accordingly, we use the average array SNR, defined as
\[\text{SNR [dB]} = 10 \log_{10} \frac{\sum_{l=1}^L E_l}{L E_w},\]
where $E_l=|R_l|^2$ is the energy of the received signal at the $l$-th hydrophone and $E_w$ is the noise power spectral density at the frequency $f=130$ Hz.\par
We evaluate models based on MAE and probability of credible localization (PCL) defined as
\[\text{PCL} = 100 \times \frac{\sum_{i=1}^{N_{test}} \mathbf{1}_{|d_i - \hat{d}_i| \leq 0.1 d_i}}{N_{test}}, \]
where $d_i$ is the true distance of the $i$-th sample and $\hat{d}_i = 100 \arg\max \hat{\mathbf{y}}_i + D_{\min} $ is the estimated distance. To train the network, we use $\sigma=5$ for label softening, a learning rate of $10^{-4}$ for the Adam optimizer, and a batch size of $128$. Furthermore, we randomly split the data into $85 \%$ training and $15 \%$ validation sets. Also, to prevent over-fitting, we multiply the learning rate by $0.1$ whenever there is no reduction in the validation error after $75$ iterations, and we stop training if there is no reduction in the validation error after $125$ iterations. The results are averaged over $100$ noise realizations.\par

\begin{figure}[htb]
\centering
\begin{subfigure}[t]{0.35\textwidth}
\centering
    \includegraphics[width= \linewidth]{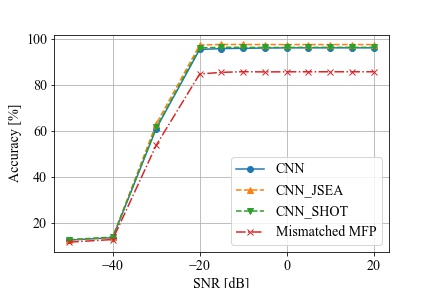}
    \caption{PCL}
    \label{fig:mae_snr}
\end{subfigure}
~
\begin{subfigure}[t]{0.35\textwidth}
\centering
    \includegraphics[width= \linewidth]{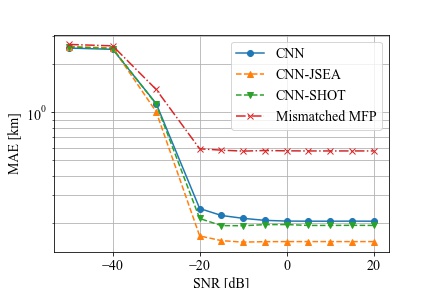}
    \caption{MAE}
    \label{fig:pcl_snr}
\end{subfigure}

\caption{Performance under different SNR values, with KAM11 noise and $\Delta c = 0.1$ m/s. }
\label{fig:snr_performance}
\end{figure}

\begin{figure}[htb]
\centering
\begin{subfigure}[t]{0.35\textwidth}
\centering
    \includegraphics[width= \linewidth]{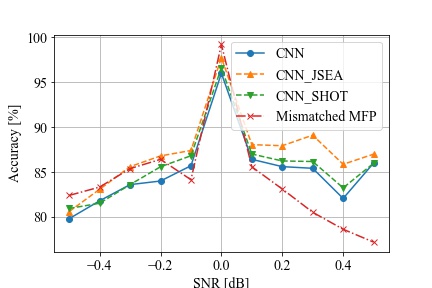}
    \caption{PCL under small mismatch in SSP.}
    \label{fig:mae_ssp_tiny}
\end{subfigure}
~
\begin{subfigure}[t]{0.35\textwidth}
\centering
    \includegraphics[width= \linewidth]{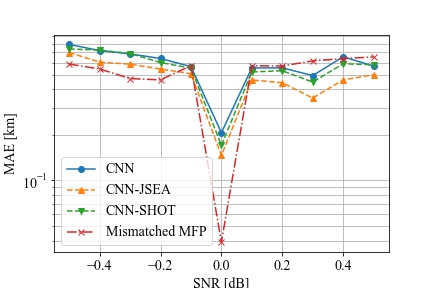}
    \caption{MAE under small mismatch in SSP.}
    \label{fig:pcl_ssp_tiny}
\end{subfigure}
\caption{Performance under different amounts of SSP mismatch at an SNR of $0$ dB. Observe the sharp deterioration of the MFP performance for the small mismatches.}
\label{fig:ssp_performance}
\end{figure}

The CNN-SHOT uses \eqref{eq:SHOT_loss} for adaptation, while CNN-JSEA uses \eqref{eq:JSEA_loss}. For both adaptation methods, we use $\mu_{DA} = 5\times10^{-6}$, and $\beta = 1$ in \eqref{eq:SHOT_loss} and we use $\delta=500$ m in \eqref{eq:pseudo}. Figure \ref{fig:snr_performance} shows the superiority of the JSEA method when the only difference between the training and testing environments is noise. In addition, according to Figs. \ref{fig:mae_ssp_tiny} and \ref{fig:pcl_ssp_tiny}, a small perturbation in the sound speed profile (along with mismatch in noise statistics) results in a significant mismatch in the arrival structure, as the MFP's performance deteriorates quickly. Nevertheless, JSEA can significantly enhance the MAE and PCL of the CNN.

% Figure \ref{fig:DA_InfoMax} shows how the MAE and average output entropy change as the adaptation progresses. Observe that as the entropy increases and approaches its limit of $\ln(82)$, the MAE decreases, although not monotonically.

% \begin{figure}[ht]
% \centering
%     \includesvg[width= \linewidth]{figures/DA_improvement_annotated.svg}
%     \caption{{\color{red} update this figure} Improvement in the localization performance during the adaptation. This figure shows only the samples whose estimation has been changed after the adaptation. The algorithm improves the performance on a collection of points as opposed to improving the performance on each individual sample.}
%     \label{fig:DA_improvement}
% \end{figure}

% \begin{figure}[htb]
% \centering
%     \includesvg[width= 0.8\linewidth]{figures/DA_InfoMax.svg}
%     \caption{Change in the MAE and outputs average entropy during the adaptation for KAM11 noise and $\Delta c = 0.1$ m/s. {\color{red} remove this figure}}
%     \label{fig:DA_InfoMax}
% \end{figure}

\section{Conclusion} \label{sec:conclusion}
By using classification for localization, the model outputs reveal the confident samples that we use to make an estimation based on the source power. We use this estimation to generate pseudo-labels for uncertain samples. We showed that while SHOT can slightly improve the performance by encouraging the network to make its predictions less uncertain on the samples similar to the training set, JSEA can significantly improve the UWA localization that almost always suffers in the presence of environmental mismatch. With more prior knowledge about domain shifts between the training and testing environments and source location distribution, one may be able to improve performance using more effective methods for selecting $\sss$ even without access to training data. However, the performance improvement will be limited in the absence of such knowledge.  \par

% \clearpage

% \bibliographystyle{IEEEtran}
% \bibliography{myrefs}

\end{document}